\documentclass[10pt,english]{article}
\usepackage[xdvi,dvips]{graphicx}
\begin{document}

\title{The electrostatic field of a point charge and an electrical dipole
in the presence of a conducting sphere}
\author{F C Santos
\footnote{e-mail: filadelf@if.ufrj.br} and A C Tort
\footnote{e-mail: tort@if.ufrj.br.}\\
 Departamento de F\'{\i}sica Te\'{o}rica - Instituto de F\'{\i}sica
\\
Universidade Federal do Rio de Janeiro\\
Caixa Postal 68.528; CEP 21941-972 Rio de Janeiro, Brazil}
\maketitle
\begin{abstract}
We evaluate the electrostatic potential and the electrostatic field 
created by a point charge and an arbitrarly oriented
electrical dipole placed near a grounded perfectly
conducting sphere. Induced surface charge distributions and possible variants of the problem are also discussed.\\
\newline \textsc{PACS} numbers: 03.50.De \\
\newline \emph{Key words}: Physics Education; Classical Electrodynamics; Electrostatics; Image Method \\
\vfill
\end{abstract}
\section{Introduction}
Several methods have been devised in order to solve electrostatic problems. Among them, the image method stands out due to its relative simplicity and strong physical appeal. Essentially, the image method leans on the uniqueness theorems of electrostatics which allow us, for instance, to replace a given conductor by a system of charges capable of reproducing the boundary conditions on
the surface of the conductor. This replacement must be done outside the physical region so that Poisson equation remains unchanged. It can be shown that it is always
possible to obtain the set of auxiliary charges necessary to solve the given electrostatic boundary problem \cite{Jackson98}. The classical example of the application of this method is the problem of a point charge at a distance $D$ from a grounded perfectly conducting infinite plate . In this case,
the image is another point charge with reversed algebraic sign at a distance $D$ below the mirror surface (the conducting plane). The electrostatic potential, and therefore the electric field, in the physical region is the sum of the potentials of the point charge and its image.
From the knowledge of the potential we can also calculate the distribuition of the electric charge induced on the plane. A more challenging problem is the one of the point
charge between two grounded conducting planes. The number of images in this case is infinite and they are distributed exactly as the specular images of a pointlike light source
placed between two parallel plane mirrors. For these two examples, the image method fully deserves its name. Notice, however, that for most electrostatic boundary problems the
images do not correspond to those obtained in the framework of geometrical optics. For example, in the problem of a point charge in the presence of a grounded perfectly
conducting sphere, also a well known example of an electrostatic problem, the latter does not work as mirror. Of course, this does not diminish the physical appeal of
the method.

In this paper we rediscuss and solve in detail another example of the collection of problems that can be solved by the image method: The determination of the electric field of an ideal static dipole with an \emph{arbitrary} spatial orientation in the presence of a grounded conducting sphere. A somewhat slightly different version of this problem is proposed in Batygin and Topygin's problem book on electrodynamics \cite{Batygin} where the student is asked to determine the system of images which is equivalent to the induced charges, the interaction energy between the dipole and the sphere, and the force and torque on the dipole. Only the answers are provided by those authors.
The determination of the electrostatic potential field of a grounded sphere and an electric dipole for the particular case where the dipole is colinear with the radius vector of the sphere was proposed in \cite{Russians} and a briefly sketched solution for the potential was also offered. To the electrical dipole pointing to an arbitrary spatial direction we also add a point charge. This configuration is sufficiently general as to allow the consideration of several special situations that can be compared with known results. As mentioned above, our aim here will be the enlargement of the previous discussions mentioned above and in what follows, we will employ the image method to obtain the potential, the electric field and the induced surface charge on the conduction sphere. We will also reobtain the electrostatic energy of the system as well as the force and torque on the dipole. We believe that a complete discussion of this electrostatic problem can be useful for undergraduate and graduate students as well as for their instructors. Also some variants of the problem can be proposed. We also briefly discuss some of these variants. S.I. units will be used.
\section{The electrostatic problem and its solution}
Let us begin by enunciating more precisely the problem we initially want to solve: Given a point charge $q$ and an arbitrarly
oriented ideal dipole $\mathbf{p}$, both placed at the same point of the space and at known distance from a grounded,
perfectly conducting sphere of radius $R$, the center of which coincides with the origen of the coordinate system,
find the electric potential and field at a point $P$ in the region exterior to the sphere, and find also the induced
surface charge density $\sigma_e$ on the conducting sphere.

In order to solve this problem, it is convenient to start by recalling that for a point charge placed at a distance
$R_1$ from the center of the sphere an image charge is formed at a distance $R_2$ on the line joining the centre
of the sphere to the original point charge. 
%
%
These distances satisfy the geometrical relation
\begin{equation}\label{geometry 1}
    R_1R_2=R^2.
\end{equation}
It seems natural then to place the image dipole at the same point below the spherical surface where the image charge is
located. With the convention that the indices $1$ and $2$ denote the real sources and their images, respectively, the
electrostatic potential at a point $P$ exterior to the sphere is tentatively written as
\begin{equation}\label{ScalarPotential}
    V\left(P\right)=\frac{\mathbf{p}_1\cdot\mathbf{r}_1}{4\pi\epsilon_0r_1^3}
    +\frac{\mathbf{p}_2\cdot\mathbf{r}_2}{4\pi\epsilon_0r_2^3}+\frac{q_1}{4\pi\epsilon_0r_1}
    +\frac{q_2}{4\pi\epsilon_0r_2},
\end{equation}
where $\mathbf{r}_1$ and $\mathbf{r}_2$ are position vectors linking the sources and their images to the
observation point $P$, respectively. The relevant boundary condition is given by
\begin{equation}\label{bc1}
V\left(M\right)=0,\;\;\;\;
\end{equation}
where $M$ denotes an arbitrary point on the surface of the conductor. Therefore, on the surface of the
conductor we write
\begin{equation}\label{bc2}
    \frac{\mathbf{p}_1\cdot\mathbf{r}_1}{r_1^3}
    +\frac{\mathbf{p}_2\cdot\mathbf{r}_2}{r_2^3}
    +\frac{q_1}{r_1}+\frac{q_2}{r_2}=0.
\end{equation}
Now we define new variables according to
\begin{equation}\label{newvariables1}
    \mathbf{R}=\mathbf{r}_1+\mathbf{R}_1,
\end{equation}
and
\begin{equation}\label{newvariables2}
    \mathbf{R}=\mathbf{r}_2+\mathbf{R}_2,
\end{equation}
where $\mathbf{R}$ is the radius vector linking the origin
to a point on the spherical surface and
$\mathbf{r}_1$ and $\mathbf{r}_2$ are now vectors
linking the dipole and its image to this same point on the spherical surface.
As in the problem of a point charge in the presence of a conducting sphere, for points on the spherical surface the
geometrical relation 
\begin{equation}\label{}
\frac{r_1}{r_2}=k,
\end{equation}
where $r_i\equiv\|\mathbf{r}_i\|$, and $k$ is a constant, holds, and we can rewrite Eq.
(\ref{bc2}) as
\begin{equation}\label{}
\left[\frac{\mathbf{p}_1}{k^3}+\mathbf{p}_2-2\left(\frac{q_1}{k}+q_2\right)\mathbf{R}_2\right]
    \cdot\mathbf{R}-\frac{\mathbf{p}_1\cdot\mathbf{R}_1}{k^3}
    -\mathbf{p}_2\cdot\mathbf{R}_2+\left(\frac{q_1}{k}+q_2\right)
\left(R^2+R_2^2\right)=0,
\end{equation}
where we have also made use of Eqs. (\ref{newvariables1}) and (\ref{newvariables2}). Since this relation holds for an
arbitrary point on the conducting surface, i.e., for arbitrary $\mathbf{R}$, we must have
\begin{equation}\label{firsteq}
\frac{\mathbf{p}_1}{k^3}+\mathbf{p}_2-2\left(\frac{q_1}{k}+q_2\right)\mathbf{R}_2=0,
\end{equation}
and
\begin{equation}\label{secondeq}
\frac{\mathbf{p}_1\cdot\mathbf{R}_1}{k^3}
    +\mathbf{p}_2\cdot\mathbf{R}_2-\left(\frac{q_1}{k}+q_2\right)\left(R^2+R_2^2\right)=0.
\end{equation}
We can solve this system for the unknown quantities $q_2$ and $\mathbf{p}_2$ as functions of the known sources $q_1$ and $\mathbf{p}_1$ in a quick way by taking the dot product of the first equation with $\mathbf{R}_2$. From the
result we subtract the second equation to obtain
\begin{equation}\label{}
    \frac{\mathbf{p}_1\cdot\left(\mathbf{R}_2-\mathbf{R}_1\right)}{k^3}
    +\left(\frac{q_1}{k}+q_2\right)\left(R^2-R^2_2\right)=0,
\end{equation}
or, solving for $q_2$
\begin{equation}\label{imagecharge1}
    q_2=-\frac{\mathbf{p}_1\cdot
    \left(\mathbf{R}_2-\mathbf{R}_1\right)}{k^3\left(R^2-R_2^2\right)}-\frac{q_1}{k}.
\end{equation}
Making use of the geometrical relation given by Eq. (\ref{geometry 1}) (keep in mind that we are working on the surface of the sphere) and noticing that $\mathbf{R}_1=R_1\hat\mathbf{R}_1$ and $\mathbf{R}_2=R_2\hat\mathbf{R}_1$, where
$\hat\mathbf{R}_1$ is a fixed unit radial vector along the direction of $\mathbf{R}_1$, we can write Eq. (\ref{imagecharge1}) as
\begin{equation}\label{imagecharge2}
    q_2=\frac{\mathbf{p}_1\cdot\mathbf{R}_1}{k^3R^2}-\frac{q_1}{k}.
\end{equation}
On the other hand, we can take advantage from the fact $\mathbf{R}$ is arbitrary and choose it in such a way that
$\mathbf{R}$ becomes parallel to $\mathbf{R}_1$ and $\mathbf{R}_2$ so that $k$ can be easily calculated
\begin{equation}\label{k}
    k=\frac{R_1-R}{R-R_2}=\frac{R_1}{R},
\end{equation}
where we have made use of Eq. (\ref{geometry 1}). Then taking this result into Eq. (\ref{imagecharge2}) we finally obtain
for the image point charge the result
\begin{equation}\label{imagecharge3}
    q_2=\frac{R\mathbf{p}_1\cdot\mathbf{R_1}}{R_1^3}-\frac{R}{R_1}q_1.
\end{equation}
Equation (\ref{imagecharge3}) fixes the image charge $q_2$ completely. Remark that for the particular configuration where the dipole is perpendicular to its position vector relative to the centre of the sphere, $\mathbf{R_1}$, or when $\mathbf{p}_1=0$, the image charge is the one of the problem of the grounded sphere and a point charge \cite{Jackson98}.

Taking Eq. (\ref{imagecharge3}) into Eq. (\ref{firsteq}), writing $\mathbf{R}_2=R_2\mathbf{\hat R}_1$ and making use again of Eq. (\ref{geometry 1}) to eliminate $R_2$ we
obtain
\begin{equation}\label{induceddipole}
\mathbf{p}_2=-\frac{R^3}{R_1^3}\left[\mathbf{p}_1-2\left(\mathbf{p}_1
\cdot\hat\mathbf{R}_1\right)\hat\mathbf{R}_1\right]\,.
\end{equation}
Equation (\ref{induceddipole}) fixes the direction and magnitude of the image dipole $\mathbf{p}_2$ completely. 
By making use of the relation $\mathbf{a}\times\left(\mathbf{b}\times\mathbf{c}\right)=\left(\mathbf{a}\cdot\mathbf{c}\right)\mathbf{b}-\left(\mathbf{a}\cdot\mathbf{b}\right)\mathbf{c}$, we can also write
\begin{equation}\label{induceddipole2}
\mathbf{p}_2=\frac{R^3}{R_1^3}\left[\mathbf{p}_1+2\left(\mathbf{p}_1
\times\hat\mathbf{R}_1\right)\times\hat\mathbf{R}_1\right]\,.
\end{equation}
The second term on the lhs of Eq. (induceddipole2) determines the transverse component of the image dipole with respect to the position vector $\mathbf{R}_1$ of the real dipole.

\begin{figure}[h!]
\begin{center}
\vskip 8cm
\includegraphics[width=10 cm]{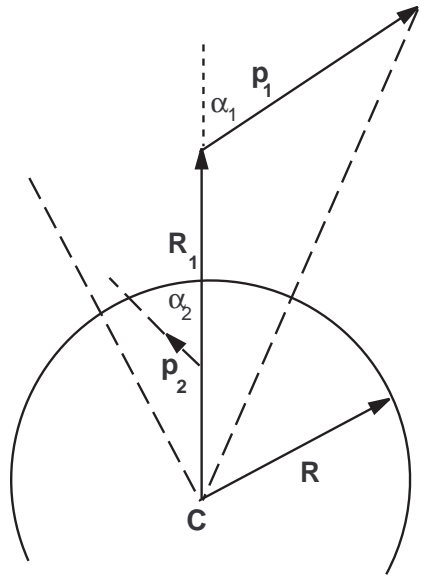}\\
\caption{The angle between $\mathbf{R}_1$ and the image dipole is the same as  angle between $\mathbf{R}_1$ and the real dipole. The magnitude of the image dipole is reduced by the factor $R^3/R_1^3$ with respect to the magnitude of the real dipole. }
\label{idipolefig2}
\end{center}
\end{figure}

From Eq. (\ref{induceddipole}) we can easily show that the magnitude of the dipole and that of its image are related by
\begin{equation}
\|\mathbf{p}_2\mid=\frac{R^3}{R_1^3}\|\mathbf{p}_1\|\,.
\end{equation}
Moreover, since 
\begin{equation}
\mathbf{p}_2\cdot\mathbf{\hat R}_1= \frac{R^3}{R_1^3}\mathbf{p}_1\cdot\mathbf{\hat R}_1\,,
\end{equation}
it also easily seen that the angle between the direction determined by $\mathbf{R}_1$ and the image dipole is the same as angle between this same direction and the real dipole. Therefore, the radial projection of image dipole and that of the real dipole are positive and add up, see Figure \ref{idipolefig2}. 
%
%

Taking Eqs. (\ref{imagecharge3}) and (\ref{induceddipole}) into Eq. (\ref{ScalarPotential}) and recalling that $\mathbf{r}=\mathbf{r}_1+\mathbf{R}_1=\mathbf{r}_2+\mathbf{R}_2$, we obtain for the electrostatic potential at an observation point $P\left(\mathbf{r}\right)$ not on the surface of the sphere the expression
%
%
\begin{eqnarray}
4\pi\epsilon_0V\left(\mathbf{r}\right)&=&\frac{\mathbf{p}_1\cdot\left(\mathbf{r}-\mathbf{R}_1\right)}{\|\mathbf{r}-\mathbf{R}_1\|^3}-\frac{R^3}{R_1^3}
\frac{\left[\mathbf{p}_1-2\left(\mathbf{p}_1\cdot\hat\mathbf{R}_1\right)\hat\mathbf{R}_1\right]\cdot\left(\mathbf{r}-\frac{R^2}{R_1}\hat\mathbf{R}_1\right)}{\|\mathbf{r}-\frac{R^2}{R_1}\hat\mathbf{R}_1\|^3}\nonumber \\
&+&\frac{R}{R_1^3}\frac{\mathbf{p}_1\cdot\mathbf{R}_1}{\|\mathbf{r}-\frac{R^2}{R_1}\hat\mathbf{R}_1\|}-\frac{R}{R_1}\frac{q_1}{\|\mathbf{r}-\frac{R^2}{R_1}\hat\mathbf{R}_1\|}+\frac{q_1}{\|\mathbf{r}-\mathbf{R}_1\|}.
\end{eqnarray}
The electric field is minus the gradient of this expression and a straightforward calculation yields
%
\begin{eqnarray}\label{totalE}
4\pi\epsilon_0\mathbf{E}\left(\mathbf{r}\right)&=&-\frac{\mathbf{p}_1}{\|\mathbf{r}-\mathbf{R}_1\|^3}+\frac{3\mathbf{p}_1\cdot\left(\mathbf{r}-\mathbf{R}_1\right)\left(\mathbf{r}-\mathbf{R}_1\right)}{\|\mathbf{r}-\mathbf{R}_1\|^5}+\frac{R^3}{R_1^3}\frac{\left[\mathbf{p}_1-2\left(\mathbf{p}_1\cdot\hat\mathbf{R}_1\right)\hat\mathbf{R}_1\right]}
{\|\mathbf{r}-\frac{R^2}{R_1}\hat\mathbf{R}_1\|^3}\nonumber \\
&-&3\frac{R^3}{R_1^3}\frac{\left[\mathbf{p}_1-2\left(\mathbf{p}_1\cdot\hat\mathbf{R}_1\right)\hat\mathbf{R}_1\right]\cdot
\left(\mathbf{r}-\frac{R^2}{R_1}\hat\mathbf{R}_1\right)\left(\mathbf{r}-\frac{R^2}{R_1}\hat\mathbf{R}_1\right)}{\|\mathbf{r}-\frac{R^2}{R_1}\hat\mathbf{R}_1\|^5}+
\frac{R}{R_1^3}\frac{\mathbf{p}_1\cdot\mathbf{R}_1\left( \mathbf{r}-\frac{R^2}{R_1}\hat\mathbf{R}_1\right)}{\|\mathbf{r}-\frac{R^2}{R_1}\hat\mathbf{R}_1\|^3}\nonumber \\
&-&\frac{Rq_1\left(\mathbf{r}-\frac{R^2}{R_1}\hat\mathbf{R}_1\right)}{R_1\|\mathbf{r}-\frac{R^2}{R_1}\hat\mathbf{R}_1\|^3} + \frac{q_1\left(\mathbf{r}-\mathbf{R}_1\right)}{\|\mathbf{r}-\mathbf{R}_1\|^3}.
\end{eqnarray}
A simple calculation -- it suffices to consider only terms proportional to $\mathbf{r}$ --  shows that the electric field on the
surface of the sphere, $\mathbf{r}=\mathbf{R}$, is given by
\begin{eqnarray}
   4\pi\epsilon_0\mathbf{E}\left(\mathbf{R}\right)&=& \frac{3}{R^3}\frac{\mathbf{p}_1\cdot\left(\hat\mathbf{r}-\frac{R_1}{R}\hat\mathbf{R}_1\right)\hat\mathbf{R}}{\|\hat\mathbf{r}-\frac{R_1}{R}\hat\mathbf{R}_1\|^5}-\frac{3}{R_1^3}\frac{\left[\mathbf{p}_1-2\left( \mathbf{p}_1\cdot\hat\mathbf{r}_1\right)\hat\mathbf{R}_1 \right]\cdot\left(\hat\mathbf{r}-\frac{R}{R_1}\hat\mathbf{R}_1 \right)\hat\mathbf{R}}{\|\hat\mathbf{r}-\frac{R}{R_1}\hat\mathbf{R}_1\|^5}+\frac{1}{R_1^2R}\frac{\left(\mathbf{p}_1\cdot\hat\mathbf{R}_1\right)\hat\mathbf{R}}{\|\hat\mathbf{r}-\frac{R}{R_1}\hat\mathbf{r}_1\|^3}\nonumber \\
&-&\frac{q_1}{R_1\,R}\frac{\hat\mathbf{R}}{\|\hat\mathbf{R}-\frac{R}{R_1}\hat\mathbf{R}_1\|^3}+\frac{q_1}{R^2}\frac{\hat\mathbf{R}}{\|\hat\mathbf{R}-\frac{R_1}{R}\hat\mathbf{R}_1\|^3}\,.
\end{eqnarray}
The induced superficial charge density on the sphere is given by $\sigma=\epsilon_0\hat\mathbf{R}\cdot\mathbf{E}\left(\mathbf{R}\right)$, therefore, the general expression for the induced charge density is up to the
constant $\epsilon_0$, the expression above with  the unit
radial vector $\hat\mathbf{R}$ omitted.
\section{The electric dipole in the presence of
a grounded conducting sphere}
Let us consider the special configuration formed by the electric dipole pointing at an arbitrary direction and a grounded conducting sphere \cite{Batygin, Russians}. To obtain the corresponding electrostatic potential, the electric field and the induced
charge on the sphere we set $q_1=0$ in the previous equations. Notice that in this case, for an
arbitrary direction, besides the image dipole $\mathbf{p}_2$, we have also, as remarked before, an image
charge $q_2$ which depends on the relative orientation between $\mathbf{p}_1$ and $\mathbf{R}_1$. Only if
$\mathbf{p}_1$ and $\mathbf{R}_1$ are mutually perpendicular the image point charge will be zero and the
image dipole will be scaled down by the factor $R^3/R_1^3$ and its direction will be opposite to that of the real
dipole. 

In order to evaluate the electrostatic energy stored in this configuration we must renormalise the electric field. This means to subtract from the total electric field as given by Eq. (\ref{totalE}) the contribution of the real dipole, i.e., the first two terms on the lhs of Eq. (\ref{totalE}). Then the electrostatic energy can be calculated from the formula
\begin{equation}
U=-\frac{1}{2}\mathbf{p}_1\cdot\mathbf{E}_{\mbox{\tiny ren}}\left(\mathbf{R}_1\right)\,,
\end{equation}
where $\mathbf{E}_{\mbox{\tiny ren}}\left(\mathbf{R}_1\right)$ is given by
\begin{eqnarray}
4\pi\epsilon_0\mathbf{E}_{\mbox{\tiny ren}}\left(\mathbf{R}_1\right)&=&\frac{R^3}{R_1^3}
\frac{\left[\mathbf{p}_1-2\left( \mathbf{p}_1\cdot\hat\mathbf{R}_1\right)\hat\mathbf{R}_1 \right]}{\|\mathbf{R}-\frac{R^2}{R_1}\hat\mathbf{R}_1\|^3}
-3\frac{R^3}{R_1^3}
\frac{\left[\mathbf{p}_1-2\left( \mathbf{p}_1\cdot\hat\mathbf{R}_1\right)\hat\mathbf{R}_1 \right]\cdot\left(\mathbf{R}-\frac{R^2}{R_1}\hat\mathbf{R}_1\right)\cdot\left(\mathbf{R}-\frac{R^2}{R_1}\hat\mathbf{R}_1\right)}{\|\mathbf{R}-\frac{R^2}{R_1}\hat\mathbf{R}_1\|^5}\nonumber \\
&+&\frac{R}{R_1^3}\frac{\mathbf{p}_1\cdot\mathbf{R}_1\left(\mathbf{R}_1-\frac{R^2}{R_1}\hat\mathbf{R}_1\right)}{\|\mathbf{R}_1-\frac{R^2}{R_1}\hat\mathbf{R}_1\|^3}\,.
\end{eqnarray}
A straightforward calculation yields for the electrostatic energy of the configuration the formula
\begin{equation}
U=-\frac{1}{2}\frac{R\left[R_1^2\left(\mathbf{p}_1\cdot\hat\mathbf{R}_1\right)^2 +R^2\mathbf{p}_1^2\right]}{4\pi\epsilon_0\left(R_1^2-R^2\right)^3}\,
\end{equation}
which is in perfect agreement with \cite{Batygin}\footnote{In order to compare our results with those of Ref. \cite{Batygin} we must set $\mathbf{R}_1=R_1\mathbf{\hat z}$, $\mathbf{p}_1= \|\mathbf{p}_1\|\left(\sin\alpha\,\mathbf{\hat x}+\cos\alpha\,\mathbf{\hat z}\right)$, where $\alpha$ is the angle between $\mathbf{R}_1$ and $\mathbf{p}_1$. Then it is readily seen that both results agree. The same hold for the induced image charge $q_2$. } 

The force on the real dipole can be calculated by taking the gradient of the electrostatic energy
\begin{equation}
\mathbf{F}_1=-\mathbf{\nabla}\,U=-\frac{d\,U}{d\,R_1}\hat\mathbf{R}_1\,.
\end{equation}
The result  is
\begin{equation}
\mathbf{F}_1=-\frac{RR_1}{4\pi\epsilon_0\left(R_1^2-R^2\right)^4}\left[\left(2R_1^2+R^2\right)\left(\mathbf{p}_1\cdot\hat\mathbf{R}_1\right)^2+3R^2\mathbf{p}_1^2\right]\hat\mathbf{R}_1\,,
\end{equation}
also in agreement with \cite{Batygin}.

To evaluate the torque on the real dipole we must consider the derivative of the eletrostatic energy with respect to the angle between $\mathbf{p}_1$ and the unit vector $\hat\mathbf{R_1}$. If we denote this angle by $\alpha$ then
\begin{eqnarray}
\tau&=&-\frac{\partial\,U}{\partial\,\alpha}\nonumber \\
&=&-\frac{R\,R_1^2}{2}
\frac{\|\mathbf{p}_1\|^2\,\sin\left(2\alpha\right)}{4\pi\epsilon_0\left(R_1^2-R^2\right)^3}\,,
\end{eqnarray}
in agreement with \cite{Batygin}.

\begin{figure}[h!]
\begin{center}
\includegraphics[width=10cm]{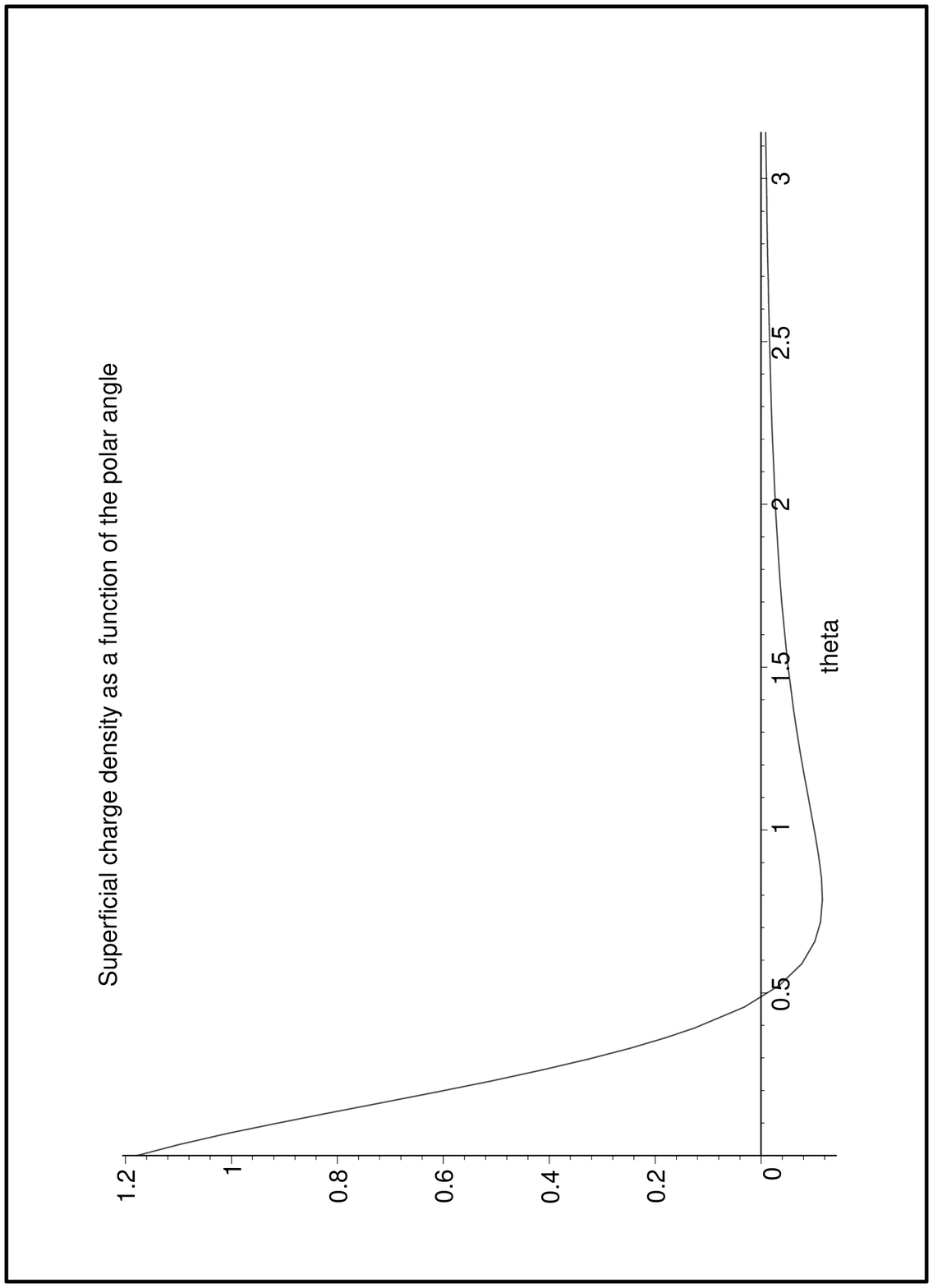}\\
\caption{Dimensionless surface charge density $\sigma^*\left(\mathbf{R}\right)$ of the grounded sphere in the configuration of Ref \cite{Batygin} as a function of the polar angle $\theta$. Here $\alpha=\pi/4$, $\phi=0$, and $R_1=2R$.}
\label{sigmastarfig1}
\end{center}
\end{figure}

\begin{figure}
\begin{center}
\includegraphics[width=10cm]{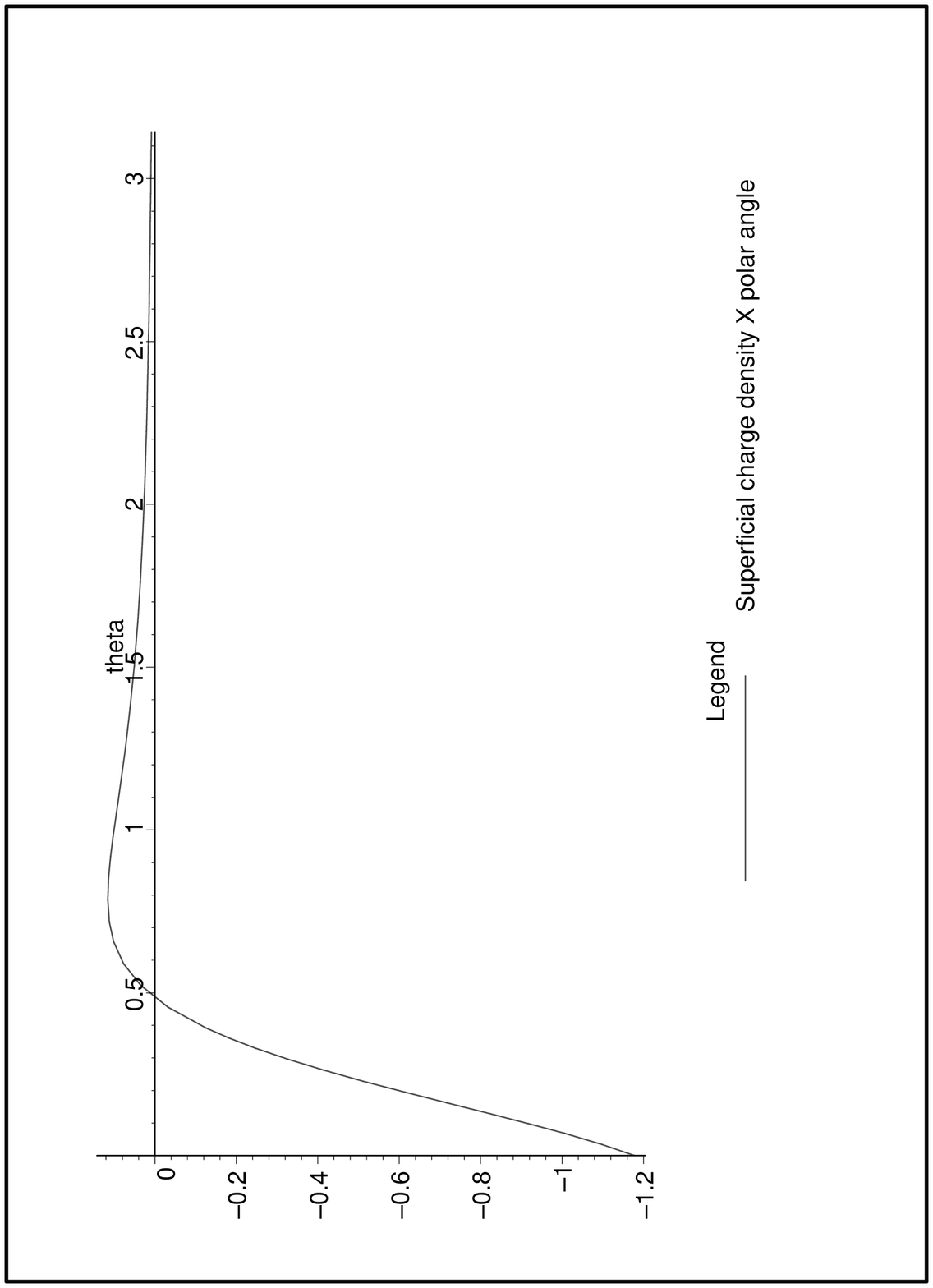}\\
\caption{Dimensionless surface charge density $\sigma^*\left(\mathbf{R}\right)$ of the grounded sphere in the configuration of Ref \cite{Batygin} as a function of the polar angle $\theta$. Here $\alpha=5\pi/4$, $\phi=0$, and $R_1=2R$.}
\label{sigmastarfig2}
\end{center}
\end{figure}

\begin{figure}
\begin{center}
\includegraphics[width=10cm]{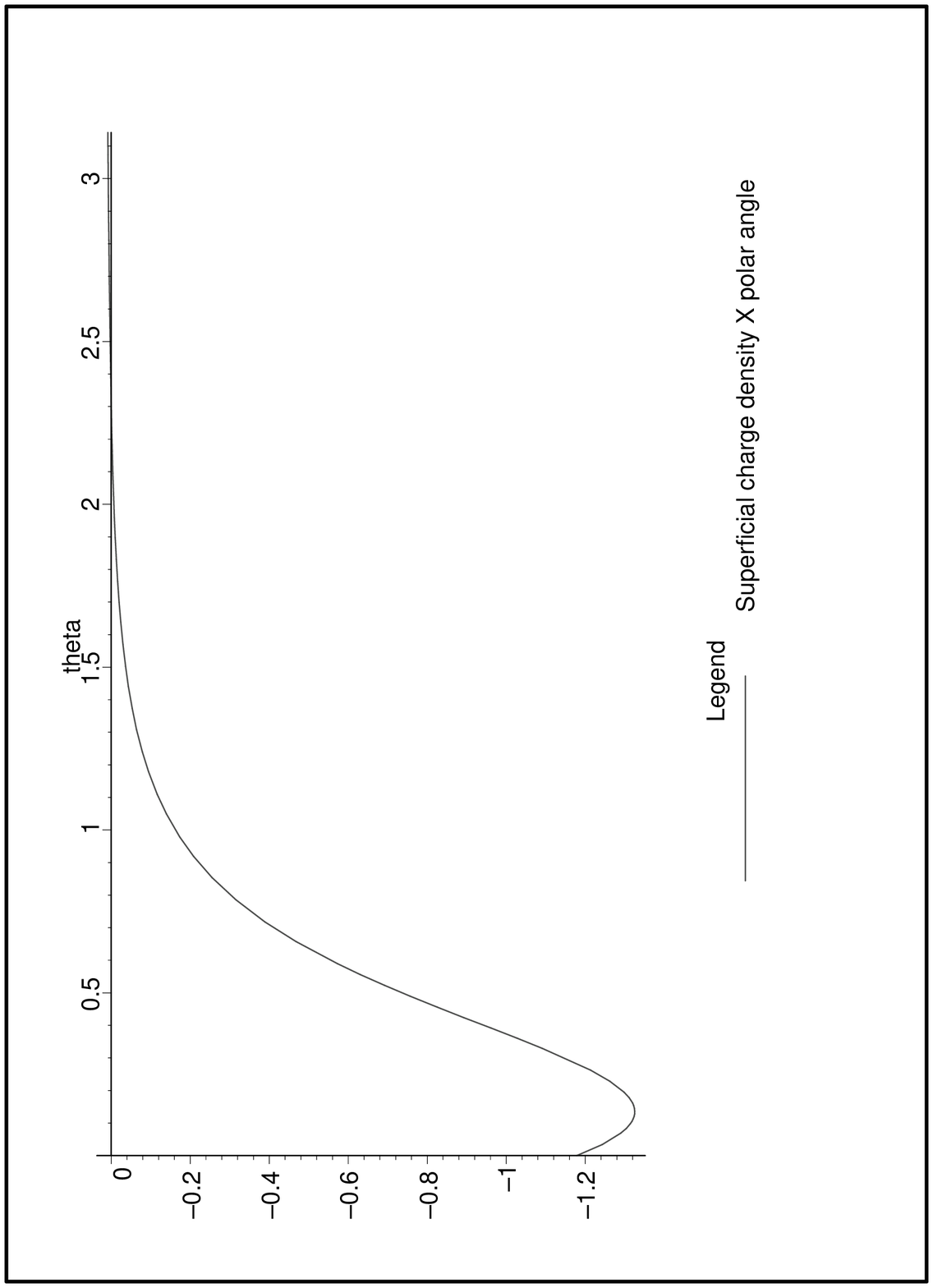}\\
\caption{Dimensionless surface charge density $\sigma^*\left(\mathbf{R}\right)$ of the grounded sphere in the configuration of Ref \cite{Batygin} as a function of the polar angle $\theta$. Here $\alpha=3\pi/4$, $\phi=0$, and $R_1=2R$.}
\label{sigmastarfig4}
\end{center}
\end{figure}

\begin{figure}
\begin{center}
\includegraphics[width=10cm]{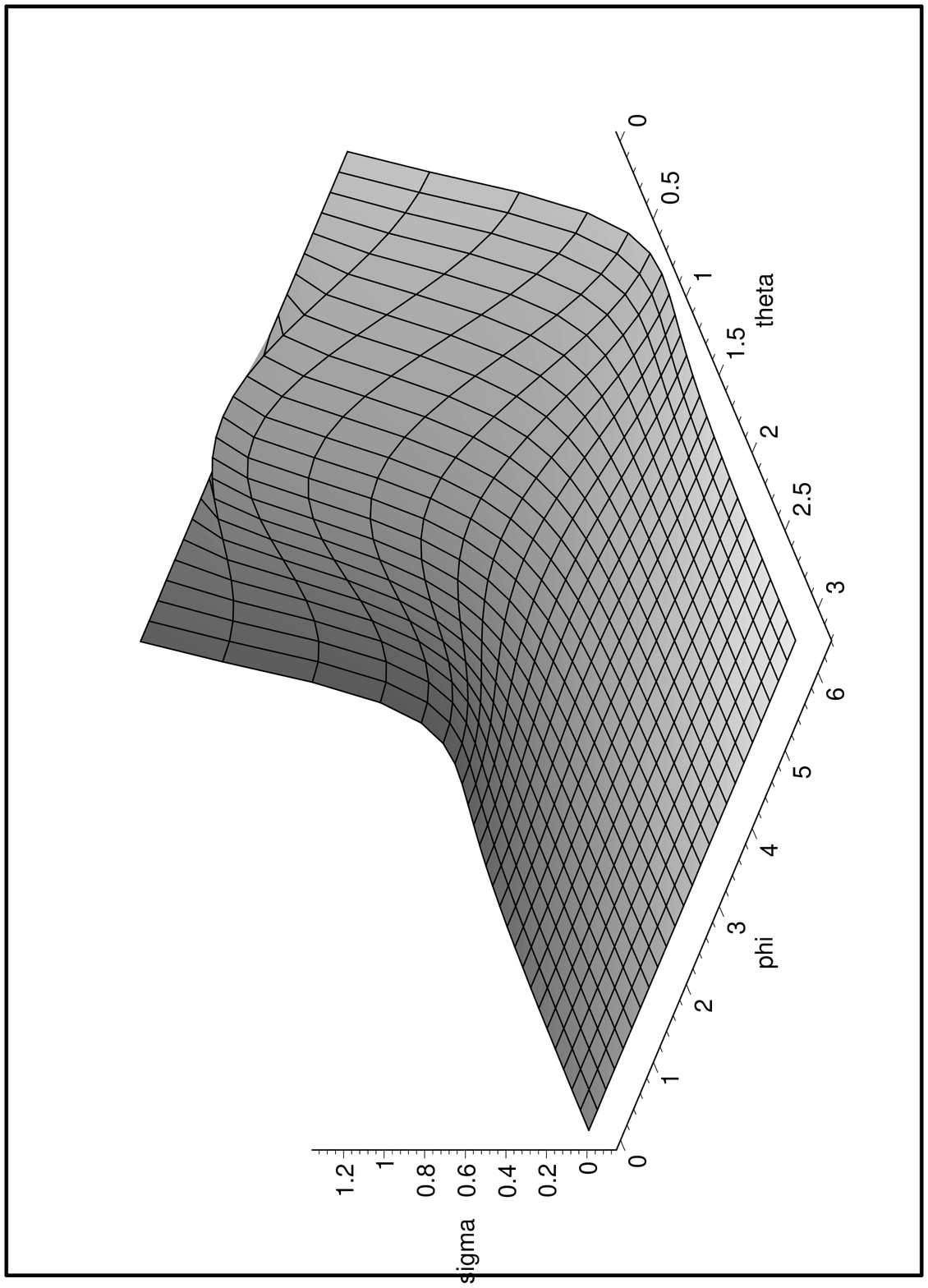}\\
\caption{Dimensionless surface charge density $\sigma^*\left(\mathbf{R}\right)$ of the grounded sphere in the configuration of Ref \cite{Batygin} as a function of the polar angle $\theta$ and the azimuthal angle $\phi$. Here $\alpha=\pi/4$, and $R_1=2R$.}
\label{sigmastarfig3B}
\end{center}
\end{figure}

The induced superficial charge distribution on the conducting sphere is
\begin{eqnarray}\label{sigma}
\sigma\left(\mathbf{R}\right)&=&\frac{3}{4\pi R^3}\frac{\mathbf{p}_1\cdot\left(\hat\mathbf{R}-\frac{R_1}{R}\hat\mathbf{R}_1\right)}{\|\hat\mathbf{R}-\frac{R_1}{R}\hat\mathbf{R}_1\|^5}-\frac{3}{4\pi R_1^3}\frac{\left[\mathbf{p}_1-2\left( \mathbf{p}_1\cdot\hat\mathbf{R}_1\right)\hat\mathbf{R}_1 \right]\cdot\left(\hat\mathbf{R}-\frac{R}{R_1}\hat\mathbf{R}_1 \right)}{\|\hat\mathbf{R}-\frac{R}{R_1}\hat\mathbf{R}_1\|^5}\nonumber\\
&+&\frac{1}{4\pi R_1^2R}\frac{\left(\mathbf{p}_1\cdot\hat\mathbf{R}_1\right)}{\|\hat\mathbf{R}-\frac{R}{R_1}\hat\mathbf{R}_1\|^3}\,.
\end{eqnarray}
%
%
In order to visualise in an easier way the induced charge density we choose the configuration of Ref. \cite{Batygin}
\begin{equation}
\mathbf{p}_1= \|\mathbf{p}_1\|\left(\sin\alpha\,\mathbf{\hat x}+\cos\alpha\,\mathbf{\hat z}\right)\,
\end{equation}
where $\alpha$ is the angle between $\mathbf{R}_1$ and $\mathbf{p}_1$. Also $\mathbf{R}_1=R_1\mathbf{\hat z}$. Then from Eq. (\ref{sigma}) the surface charge density can be cast into the expression
\begin{eqnarray}\label{sigmastar}
\sigma^*\left(\mathbf{R}\right)&=& \frac{\left[\hat\mathbf{x}\cdot\hat\mathbf{R}\sin\alpha+\left(\hat\mathbf{z}\cdot\hat\mathbf{R}-B^{-1}\right)\cos\alpha\right]}{\left(1+B^{-2}-2A^{-1}\hat\mathbf{x}\cdot\hat\mathbf{R}\right)^{5/2}}-B^3\frac{\left[\hat\mathbf{x}\cdot\hat\mathbf{R}\sin\alpha-\left(\hat\mathbf{z}\cdot\hat\mathbf{R}-B\right)\cos\alpha\right]}{\left(1+B^2-2B\hat\mathbf{x}\cdot\hat\mathbf{R}\right)^{5/2}} \nonumber \\
&+&\frac{B^2}{3}\frac{\cos\alpha}{\left(1+B^2-2B\hat\mathbf{x}\cdot\hat\mathbf{R}\right)^{3/2}}\,,
\end{eqnarray}
where we have defined $\sigma^*\left(\mathbf{R}\right):=4\pi R^3\sigma\left(\mathbf{R}\right)/3\|\mathbf{p}_1\|$, and $B:=R/R_1$; also $\hat\mathbf{x}\cdot\hat\mathbf{R}=\sin\theta\cos\phi$, and $\hat\mathbf{z}\cdot\hat\mathbf{R}=\cos\theta$. The angles $\theta$ and $\phi$ are the usual polar and azimuthal angles associated with spherical coordinates. Notice that $0\leq B\leq 1$. Also, for $R_1\gg R$ we have $B\to 0$ and $\sigma^*\left(\mathbf{R}\right)\to 0$, as it should. Equation (\ref{sigmastar}) will allow for simple plots of the dimensionless surface charge distribution as a function of the polar angle once we have chosen $\alpha$ and $\phi$. As an example we display in Fig. \ref{sigmastarfig1} the case in which we set $\alpha=\pi/4$ and $\phi=0$. More examples are shown in the subsequent Figs. \ref{sigmastarfig2}, \ref{sigmastarfig4}. Of course we can also construct three-dimensional plots such as that of Fig. \ref{sigmastarfig3B}. 
%
%
%
%
%
\section{Final remarks}
From the general solutions that we have found we can answer a number of particular questions, for example, we now can readly answer to the question \cite{Eyge1972}: A point dipole of moment $\mathbf{p}$ is a distance $d$ from the center of a grounded conducting sphere and points to the
center of it. What is the charge distribution on the sphere? Or we can answer to the question: Does it matter if the sphere is isolated or grounded \cite{Konopinski1981}? For the grounded sphere the total charge is not specified and the amount of charge on it will depend on the details
of the configuration. For an isolated charged sphere the total charge $Q$ has a fixed value and the potential on the surface of the sphere has a value $V_0$. In this case, an additional point charge $q_3$ placed at the center of the sphere will solve the problem. If the sphere is isolated
but neutral, the condition $q_3+q_2=0$ must hold.
\section*{Acknowledgments}
The authors wish to acknowledge their students for helpful comments.

%
\newpage
\section*{ FIGURES AND CAPTIONS}
\newpage
\begin{figure}
\begin{center}
\includegraphics[width=15 cm]{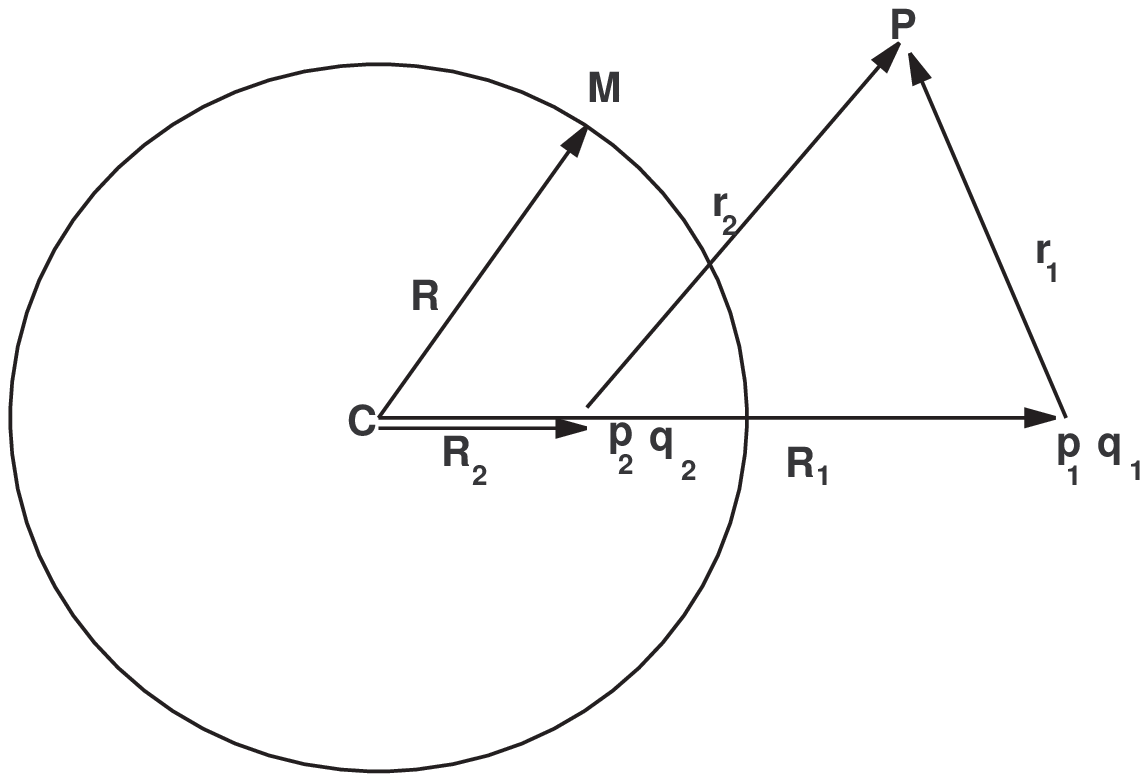}\\
\caption{Geometry for the problem of a dipole and a point charge in the presence of a grounded perfectly conducting sphere. The dipole has an arbitrary orientation in three-dimensional space.}
\label{idipolefig1}
\end{center}
\end{figure}
\newpage
\newpage
\newpage
\newpage
\newpage
\end{document}